\documentstyle[12pt]{article}
\begin{document}
\begin{flushright}
TAUP-2319-96\\
\end{flushright}
\baselineskip 24pt
\newcommand{\be}{\begin{equation}}
\newcommand{\ee}{\end{equation}}
\newcommand{\leqx}{\,\raisebox{-1.0ex}{$\stackrel{\textstyle <}
{\sim}$}\,}
\newcommand{\geqx}{\,\raisebox{-1.0ex}{$\stackrel{\textstyle >}
{\sim}$}\,}
\newcommand{\th}{\theta}
\newcommand{\va}{\varphi}
\newcommand{\de}{\Delta}
\newcommand{\x}{\tilde{x}}
\newcommand{\bl}{\hspace{-.65cm}}
\newcommand {\s}
[1] {\mid \! \! {#1} \rangle}  
\begin{center}
{\bf Some remarks on 't Hooft's S-matrix for black holes}
\\N.Itzhaki \footnote{Email Address:sanny@post.tau.ac.il}
\\Raymond and Beverly Sackler Faculty of  Exact Sciences
\\School
 of Physics and Astronomy
\\Tel Aviv University, Ramat Aviv, 69978, Israel
\end{center}
\begin{abstract}
We discuss the limitations of  't Hooft's proposal for the
 black hole S-matrix.
We find that the validity of the S-matrix implies violation 
of the semi-classical approximation at scales large compared
 to the Planck scale.
We also show that the effect of the centrifugal barrier on the
S-matrix is crucial even for large transverse distances.
\end{abstract}
\newpage

{\bf 1. Introduction}

A central question in the understanding of black holes is 
 how  the gravitational interaction influence the final state
 of  Hawking  radiation. 
Usually,  when $R\ll 1$ ($R$ is the Riemann curvature)
\footnote{in units where $G=\hbar =c=1$} the 
gravitational effect is small.
Hawking concluded therefore that strong gravitational effects 
take place only near the singularity (where  $R\approx 1$),
 far behind the horizon, so the radiation is not affected by it
\cite{haw}.
$R_{hor}\approx \frac{1}{M^2}\ll 1$,
thus as long as $M\gg 1$ the gravitational interaction is only
 via adiabatic change of the mass of the black hole.

't Hooft, however, pointed out that even for $M\gg 1$ strong
 gravitational interactions occur near the horizon due to
 the strong red-shift effect  and that these interactions are important for
 the final state of the radiation.
Moreover, he was able to suggest an S-matrix for the black hole formation
 and 
evaporation process \cite{th2} based on the classical gravitational shock
wave! \cite{th1}
In this note we present objections to the construction of the S-matrix.
It is important to emphasize that these objections are {\em not}
to 't Hooft's
S-matrix ansatz or to the claim  that the gravitational interactions
 near the horizon play
an important role in the black hole puzzle but only to the
 specific
derivation of the  S-matrix.
First let us briefly summarize 't Hooft's construction of the S-matrix
(the full details are in refs.\cite{th2,th1,th4,th3}).

{\bf 2. The S-matrix}

The gravitational field of a massless particle in Minkowski space is
described by  the line element
\be ds^2 =-du(dv+4p\ln (\x^2)\delta (u-u_0)du)+dx^2+dy^2\ee
where $\x ^2=x^2+y^2$.
The massless particle moves in the $v$ direction with constant
 $u_0$ and momentum $p$ \cite{sex}.

There are two effects of such a shock wave on null geodesics :

\bl 1-A discontinuity $\delta v$ at $u=u_0$.
The condition on $\delta v$ is 
\be \triangle (\delta v)=-16\pi p\delta (\x ).\ee
The solution of this equation is 
\be \delta v(\x )=-4p\ln (\x ^2 )+C\ee
where $C$ is an arbitrary constant.
For Schwarzchild black hole one get
\be \triangle (\delta v)-\delta v=-2\pi kp\delta(\th ),\ee
where $k$ is a numerical constant which depends on the black hole mass.
The additional term $\delta v $ is due to
 the curvature so in the
 limit of small $\th $ the solution of eq.(4) reduces to the 
solution of eq.(2) with a well-defined value of $C$. 

\bl 2-A refraction in the direction of the null geodesic.
The refraction is defined by
\be \cot \alpha +\cot \beta=\frac{4p}{\x},\ee
where $\cot \alpha=\frac{\partial y}{\partial u} $ for $u<u_0$
and $\cot \beta=\frac{\partial y}{\partial u} $ for 
$u>u_0$ (we have  assumed $x=0$ without loss of generality).

Clearly, when $p$ is small $\delta v$ is small also, when we
 describe the interaction in Kruskal coordinates.
Let us see what the interaction looks like in Schwarzchild
coordinates. Kruskal coordinate and Schwarzchild coordinate are
related by $t=2M\log (\frac{u}{v})$ and $ \rho ^2=uv$. Where
 $\rho $ is the invariant distance from the horizon where the
 outgoing particle crosses the shock wave of the ingoing
 particle.
Suppose that the Schwarzchild
energy of the ingoing particle is $E$ then
\be p_{in}=8E\frac{M}{u_0},\ee
so up to a $\log (M)$ factor we get from eq.(3),
\be \delta t\approx M\frac{\delta v}{v}\approx M^2
 \frac{E}{\rho ^2}.\ee
The interaction, Eq.(7), is strong if the state of the 
outgoing particle is orthogonal to the
state that the outgoing particle would have had in the absence
 of the ingoing particle. 
Then the information is completely emitted.
This means that
\be \delta t>\frac{1}{\omega},\ee
where the energy of the outgoing particle is $\omega $. 
From eq.(7) we find that the information is completely emitted 
if
\be \rho ^2<M^2 E\omega.\ee

More generally, one can specify the particles that produced
 the black hole
\be \s{p_{in}(\Omega)}. \ee
One can also specify the particles the black hole 
decay to
\be \s{p_{out}(\Omega^{'} )}. \ee
According to the S-matrix ansatz to a good approximation (large
transverse distances)    the properties of the
black hole are completely determined by the 
S-matrix
\be \langle p_{out}(\Omega^{'} )\mid p_{in}(\Omega )\rangle.\ee
Suppose now that we add a single light incoming particle with 
momentum  $\delta p_{in}$ and angular coordinates $\Omega^{''}$.
The in-state is now
\be \s{in}^{'}=\s{p_{in}(\Omega)+\delta p_{in}\delta
(\Omega-\Omega^{''})}\ee
The effect of the shock wave on the outgoing particles is  
\begin{eqnarray}\s{out}^{'}=\exp \left(i\int d^2\Omega ^{'}
p_{out}(\Omega^{'} )\delta v(\Omega ^{'} )\right)\s{p_{out}
(\Omega ^{'})}=\\ \nonumber
\exp \left(ik \int d^2\Omega ^{'}p_{out}
(\Omega ^{'})f(\Omega^{'} ,\Omega^{''})\delta p_{in}
(\Omega ^{''})\right)
\s{p_{out}(\Omega^{'} )}. \end{eqnarray}
where $f$ is the Green-function determined by eq.(4).
Up to an overall phase only one S-matrix agrees with eq.(14)
\be  \langle p_{out}(\Omega^{'} )\mid p_{in}
(\Omega )\rangle=
N\exp \left(ik d^2\Omega d^2\Omega ^{'}p_{out}
(\Omega )f(\Omega ,\Omega^{'}) p_{in}(\Omega^{'})\right)\ee
Clearly, this leads to  ultra violet divergences because of 
the $\ln (\x ^2)$ in eq.(3).
There should be , therefore, a transverse cut-off, resulting from 
the full Hilbert space yet to be found. As a  
result the horizon entropy should be finite.
But for large transverse distances eq.(15) is supposed
 to be a good approximation to the ultimate theory of black
 hole formation and evaporation.  

{\bf 3. Difficulties with the S-matrix}

We have two independent objections to the suggested S-matrix.
These  objections do not rest on assumptions
on the Planck scale physics but only on conventional physics.
Nevertheless, they lead to important modification of the 
 S-matrix even for {\em large} transverse distances.

\bl 1. There is a hidden assumption in eq.(14).
The hidden assumption is that  
$\s{out}^{'}$ is a state containing
 outgoing particles only.
In other words the possibility that the outgoing particle is  
dragged back into 
the horizon by the ingoing particle ($\delta v>v$) is not
 taken into account.
If this happens, then the interaction between the in-state
 and the out-state is certainly not described by eq.(15).
In order to avoid this possibility
 we must impose that $\delta v<v$, which from eq.(7) means that 
 (up to a $\log (M)$ factor)
\be \rho^2\geq ME,\ee
This limitation can be understood (up to the $\log (M) $
 factor) 
in the following way:
when a particle with energy $E$ is thrown into a black hole
a new horizon is formed at $R=2M+2E$
and the invariant distance from $R=2M+2E$ to $R=2M$ in the
 original background metric is $\approx\sqrt{ME}$.
Eqs.(9, 16) implies therefore that  the interaction region,
meaning the region where the discontinuity effect is strong
 enough so the information is fully emitted with Hawking
 radiation but it is not too strong so the outgoing particle 
is not dragged into the horizon, is
\footnote{An important 
conclusion from this equation is that the energy of the
 particles which  might contain the information is larger
 then $\approx \frac{1}{M}$ this supports the 
 suggestion that the  energy spectrum of black holes is
 quantized in a way that $M^2 =cn$ where $n$ is an integer. }
\be ME<\rho ^2 <M^2 E\omega.\ee

The fact that the interaction region is bounded from below is
disturbing and was already discussed in \cite{th3}
where it was argued that one is required to consider only 
those outgoing particles that emerge later than the time 
interval $\de t=4M\ln (M)$,
and that all other particles are completely determined by 
earlier events.
$\de t$ is the time it takes for the ingoing particles to reach
$\rho= 1$ from $\rho =M$ .
This means that the interaction region is bounded from below 
by the Planck scale.
If this were the case then one could not argue against it,
since eq.(15) is based on semi-classical arguments which are 
probably incorrect at the Planck scale any way.
But we find (eq.(16)) that the interaction region is bounded
from below not by $\approx 1$ but by $\approx \sqrt{ME}\gg 1 $.
This means that   the validity of eq.(15)
requires that  the semi-classical approximation is incorrect
 for scales much larger then $1$!
This is crucial since  the usual arguments for information loss rest on the 
assumption that the semi-classical approximation is correct
 for scales much larger then $1$. 

\bl 2. Due to  the refraction effect there is another case in
 which  $\s{out}^{'}$ does not contain only outgoing 
particles: The effect of the refraction is to bend the trajectory of
the outgoing particle.
If this happens close enough to the horizon then the trajectory of a
light signal which moved in the radial direction before it crosses the
shock wave is such that it almost reaches $R=3M$ before it 
falls back to the black hole.
As shown in the appendix only for  
\be \rho ^2>EM\frac{M}{\x}\ee
the   particle
 will not fall back into the horizon.

Since $M>\x$ it is obvious that at the interaction  region
 most of the particles (recall that for Hawking radiation 
$\omega\approx\frac{1}{M}$) that are  outgoing particles  in
 the absence of  the ingoing particle will
fall back to the horizon in the presence of the ingoing 
particle. Also, part of the Hawking photons which were 
supposed to fall back to the horizon in the absence of the
 ingoing particle (large angular momenta) will be outgoing 
particles in the presence of the ingoing particle.
Another way to illustrate this point is by using eq.(14) 
directly.
From eqs.(14, 6) and $p_{out}=8\omega \frac{M}{v}$ we find that the 
difference in the phases of
 two points $\Omega _1, \Omega _2$ due to the shock wave of an 
ingoing particle at $\Omega _0$ is  
\be p_{in}(\Omega _0 )(p_{out}( \Omega _1 )
f(\Omega _0, \Omega_1)-p_{out}( \Omega _2 )
f(\Omega _0, \Omega_2))  \approx \frac{E\omega M^2}
{\rho ^2}\log \frac{\Omega _1 -\Omega _0}
{\Omega _2 -\Omega _0 },\ee
when $ \frac{1}{M} < \Omega _i -\Omega _0 < 1$ for $i=1, 2$.
Consider  Hawking photon in the S-wave state.
After it crosses the shock wave his state is a superposition
of angular momenta.
 From eq.(19) we find that  the uncertainty in the angular 
momenta is
\be \de l\approx \frac{E\omega M^2}{\rho ^2}.\ee
The centrifugal barrier effect is such that only 
photons with
\be l<\omega M\ee
can cross it.
From eqs.(20, 21) we see  that if the collision occurs at the interaction 
region $\rho^2\approx EM$ part of the S-wave  photons will be
reflected back by the centrifugal barrier.
Furthermore, part of the 
photons which high angular momenta before they cross the shock
 wave (but not to high, $l$ of the
 order of $\de l$) will not be reflected by the 
centrifugal barrier due to the shock wave.

{\bf 4. Summary}

The suggested S-matrix (eq.(15)) considers  the  shock wave
 effects near the horizon without taking into account the
 centrifugal barrier.
 Although the   centrifugal barrier has nothing to do
 with Planckian physics we find that it plays an important 
role for the ultimate S-matrix of black hole,
so,  eq.(15) is not a good approximation to the 
 ultimate S-matrix of black hole even at large transverse
distances.
Furthermore, even without taking into account the effect of 
the centrifugal barrier on the S-matrix we find that the 
validity of the suggested S-matrix leads to violation of the 
semi-classical approximation at large scales compared to the
 Planck scale.
We suspect that any mechanism which supposed to emit the information
with Hawking radiation must necessarily violate the semi-classical
approximation at large scale.

\vspace{1.5cm}

\centerline{\bf Acknowledgment }
I am grateful to Prof. G. 't Hooft for helpful comments and criticism
and to  Prof. A.Casher and Dr. S. Massar for   fruitful discussions .
I would also like to thank  to Prof. Y. Aharonov  and 
Prof. F. Englert  for useful discussions.

\vspace{1.5cm}

{\bf Appendix}

Consider the region near the black hole horizon.
In that region the  geometry just outside the event horizon
 is described
 by the Rindler metric:
\be  d\tau ^2=dT^2-dZ^2-dX^i dX^i.\ee
In terms of Schwarzchild coordinates it is given by
\be  d\tau ^2=(\frac{dt}{4M})^2\rho ^2-d\rho ^2 -dX^i dX^i.\ee
Where $\rho $ is the invariant distance from the horizon and $t$ is 
Schwarzchild time.
The  Minkowski and Schwarzchild coordinates are related by
\[Z=\rho \cosh(\frac{t}{4M})\]

\be T=\rho \sinh(\frac{t}{4M}).\ee
Suppose that an outgoing light particle crosses the shock wave of the
ingoing particle at the coordinates
\be  x_1^{\mu}=(T=0,\;\;X=0,\;\;Y=0,\;\;Z=\rho_0)\ee
and that the ingoing particle coordinates is
\be T=0,\;\;X=0,\;\;Y=\x ,\;\;Z=\rho_0.\ee
The transverse distance between the in and outgoing particles is
$\x $.
Before the collision the outgoing particle's direction was radial
$\frac{\partial Y}{\partial Z}=0$. 
From eq.(5)
we find that after the collision with the shock wave 
the trajectory is such that 
\be \frac{\partial Y}{\partial Z}<\frac{\partial Y}{\partial U}=
\frac{4p}{\x}.\ee
Let us find the condition for the particle to fall back into the horizon.
In that case  the photon will reach the point
\be x_2^{\mu}=(T, 0, Y, Z),\ee
where $ Z^2-T^2=\rho _{0}^{2}.$
We consider null geodesic  so $(x_1^{\mu}-x_2^{\mu})^2=0$
thus
\be Y^2=2\rho _0 (Z-\rho _0).\ee
Since, $\frac{\partial Y}{\partial Z}=\frac{Y}{Z-\rho_0}$ and
 $Y<M$ we get from eq.(27)
\be \rho_0<\frac{2pM}{\x}\ee
Finally from eq.(6) we get eq.(17).

\end{document}